\documentstyle[11pt,epsfig]{article}
\begin{document}
\title{\Large\bf  Understanding masses of $c\bar{s}$ states in Regge phenomenology}

\author{\small De-Min Li \footnote{E-mail: lidm@zzu.edu.cn}, ~~Bing Ma,~~Yun-Hu Liu\\
\small   Department of Physics, Zhengzhou University, Zhengzhou,
Henan 450052, P. R. China}
\date{\today}
\maketitle
\vspace{0.5cm}

\begin{abstract}

In the framework of Regge phenomenology, masses of the charmed
states $c\bar{q}~(q=u,d,s)$ lying on the $1^3S_1$-like
trajectories are estimated. The overall agreement between our
estimated masses and the recent predictions given by modified
quark models[hep-ph/0605019, hep-ph/0608011, hep-ph/0608139] is
good. Masses of the observed charmed states $D_{s0}(2317)$,
$D_{sJ}(2860)$ and $D_{sJ}(2690)$/$D_{sJ}(2700)$ can be reasonably
reproduced in the picture of these charmed states as simple
quark-antiquark configurations. We therefore suggest the
$D_{s0}(2317)$ can be identified as the $c\bar{s}(1^3P_0)$ states.
The possible assignments of the $D_{sJ}(2860)$ and $D_{sJ}(2690)$/
$D_{sJ}(2700)$ are discussed.

\end{abstract}

\vspace{0.5cm}


\newpage

\baselineskip 24pt

\section{Introduction}
\indent \vspace*{-1cm}

Recently, discovery of the new charm-strange state $D_{s0}(2317)$
\cite{ds2317ex} generated strong interest in charmed meson
spectroscopy. The $D_{s0}(2317)$ seems an obvious candidate for
the $1^3P_0$ $c\bar{s}$ state since it is the first observed
charm-strange $0^+$ resonance. However, the observed mass of the
$D_{s0}(2317)$ is more than one hundred MeV than the constituent
quark model predictions for the $1^3P_0$ $c\bar{s}$. For example,
the measured mass of the $D_{s0}(2317)$ is $2317.3\pm 0.6$
MeV\cite{pdg}, while the prediction of the $1^3P_0$ $c\bar{s}$
state by Isgur and Godrey is 2.48 GeV\cite{p1} and that by Di
Pierro and Eichten is 2.487 GeV\cite{p2}. It is widely accepted
that the constituent quark model offers the most complete
description of hadron properties and is probably the most
successful phenomenological model of hadron structure\cite{rmp}.
Therefore, the substantially small observed mass of the
$D_{s0}(2317)$ led to many exotic interpretations on the
underlying structure of the $D_{s0}(2317)$, such as the $(DK)$
molecule, four-quark state, $D\pi$ atom or baryonium have been
proposed in the literature (For the detailed review see {\sl e.g.}
Refs.\cite{rev1,rev2,rev3}).

It should be noted that it is very important to exhaust possible
conventional $c\bar{s}$ descriptions of the $D_{s0}(2317)$ before
resorting to more exotic models, and the discrepancy between the
measured mass of the $D_{s0}(2317)$ and the quark model
predictions for the $1^3P_0$ $c\bar{s}$ mass maybe imply that
approximations to the constituent quark model are not appropriate.
 In fact, it has been pointed out by Matsuki et al.\cite{matsuki} that conventional quark models
 do not completely and consistently respect heavy quark symmetry and the $D_{s0}(2317)$ mass can be
reproduced if one treats a bound state equation appropriately.
Also, it is found by Lee et al.\cite{loop} that with the one loop
chiral corrections, the quark model for $c\bar{s}$ mesons can
naturally account for the unusually mass of the $D_{s0}(2317)$.
More recently, it is shown that a simple modification to standard
vector Coulomb plus scalar linear quark potential model maintains
good agreement with the charmonium spectrum and agrees remarkably
well with the $D$ and $D_s$ spectra\cite{newmodel1,newmodel2}.

In the present work, we shall show in Regge phenomenology, masses
of the recent observed charmed states $D_0(2290)$, $D_{s0}(2317)$,
$D_{sJ}(2860)$\cite{0607082} and
$D_{sJ}(2690)/D_{sJ}(2700)$\cite{0607082,0608031} can be
reproduced within a simple $c\bar{q}$ picture, and our estimated
masses are in good agreement with those predicted
by\cite{matsuki,newmodel1,newmodel2}. Therefore, our analysis
supports the conclusion that it is not necessary to introduce more
exotic models for understanding masses of these reported charmed
states.

\section{Regge phenomenology}
\indent \vspace*{-1cm}

Regge theory is concerned with the particle spectrum, the forces
between particles, and the high energy behavior of scattering
amplitudes\cite{regge}. One of the most distinctive features of
Regge theory is the Regge trajectory by which the mass and the
spin of a hadron are related. Knowledge of the Regge trajectories
is useful not only for spectral purpose, but also for many
non-spectral purpose. The intercepts and slopes of the Regge
trajectories are of fundamental importance in hadron
physics\cite{bas}.

 A series of recent papers by Anisovich et al.\cite{anisovich} show that meson states
fit to the quasi-linear Regge trajectories with sufficiently good
accuracy, although some suggestions that the realistic Regge
trajectories could be nonlinear exist\cite{nonlinear1,2}.

The quasi-linear Regge trajectories for a meson multiplet can be
parameterized as
\begin{equation}
J=\alpha_N(t)=\alpha_{i\bar{i^\prime}(N)}(0)+\alpha^\prime_{i\bar{i^\prime}(N)} t,
\label{trajectory}
\end{equation}
where $i$ ($\bar{i^\prime}$) refers to the quark (antiquark) flavor, $i\bar{i^\prime}(N)$ denotes
the meson $i\bar{i^\prime}$ with radial quantum number $N$ ($N=1,2,3,...$),
$t=M^2_{i\bar{i^\prime}(N)}$, $J$ and $M_{i\bar{i^\prime}(N)}$ are respectively the spin and mass
of the $i\bar{i^\prime}(N)$ meson, $\alpha_{i\bar{i^\prime}(N)}(0)$ and
$\alpha^\prime_{i\bar{i^\prime}(N)}$ are respectively the intercept and slope of the trajectory on
which the $i\bar{i^\prime}(N)$ meson lies. For a meson multiplet, the parameters for different
flavors can be related by the following relations( see Ref.\cite{dmli} and references therein):

(i) additivity of intercepts,
\begin{equation}
\alpha_{i\bar{i}(N)}(0)+\alpha_{j\bar{j}(N)}(0)=2\alpha_{i\bar{j}(N)}(0),
\label{intercept}
\end{equation}

(ii) additivity of inverse slopes,
\begin{equation}
\frac{1}{\alpha^\prime_{i\bar{i}(N)}}+\frac{1}{\alpha^\prime_{j\bar{j}(N)}}=\frac{2}{\alpha^\prime_{i\bar{j}(N)}}.
\label{slope}
\end{equation}
From (\ref{trajectory}) and (\ref{intercept}), we obtain that
\begin{eqnarray}
M^2_{i\bar{i}(N)}\alpha^\prime_{i\bar{i}(N)}+M^2_{j\bar{j}(N)}\alpha^\prime_{j\bar{j}(N)}=
2M^2_{i\bar{j}(N)}\alpha^\prime_{i\bar{j}(N)}. \label{mass-1}
\end{eqnarray}

 The main purpose of this work is to discuss whether masses of the recent observed charmed states such as the $D_0(2290)$, $D_{s0}(2317)$, $D_{sJ}(2860)$ and
 $D_{sJ}(2690)/D_{sJ}(2700)$ can be reproduced correctly in the conventional quark-antiquark picture based on Regge phenomenology.
Because the possible quantum numbers of the $D_{sJ}(2860)$ include
$0^+$,
 $1^-$, $2^+$ and $3^-$\cite{0607082}, $J^P$ of the
$D_{sJ}(2700)$ ($D_{sJ}(2690)$) is $1^-$\cite{0608031}\footnote{In
$D^0K^+$ system, BaBar observed a resonance ($D_{sJ}(2690)$) with
a mass $2688\pm 4\pm 2$ MeV and width $112\pm 7\pm 36$
MeV\cite{0607082}, Belle subsequently observed a resonance
($D_{sJ}(2670)$) with a mass $2715\pm 11^{+11}_{-14}$ MeV, width
$115\pm 20^{+36}_{-32}$ MeV and $J^P=1^-$\cite{0608031}, we regard
them compatible, and the $D_{sJ}(2690)$ and $D_{sJ}(2700)$ should
be the same one resonance with $J^P=1^{-}$.}, and the $0^+$,
$1^-$, $2^+$ and $3^-$ trajectories are parity partners\cite{2},
in the following text, we shall adopt two assumptions: (a) the
slopes of the parity partners' trajectories coincide, as proposed
by Ref.\cite{2}, and (b)
$\alpha^\prime_{i\bar{j}(N)}=\alpha^\prime_{i\bar{j}(1)}$, as
adopted by Ref.\cite{anisovich}. Under these two assumptions, the
$0^+$, $1^-$, $2^+$ and $3^-$ are the $1^3S_1$-like trajectories.
In this work, the slopes of the $1^3S_1$-like trajectories used as
input are taken from our previous work\cite{dmli} and these slopes
are shown in Table 1. In the following, $n$ denotes a $u$ or $d$
quark.

\begin{table}[htb]
\begin{center}
\begin{tabular}{ccccc}\hline
$\alpha^\prime_{n\bar{n}(1)}$& $\alpha^\prime_{n\bar{s}(1)}$
&$\alpha^\prime_{c\bar{c}(1)}$&$\alpha^\prime_{c\bar{n}(1)}$&$\alpha^\prime_{c\bar{s}(1)}$\\\hline
$0.8830$ &$0.8493$&$0.4364$ &$0.5841$&$0.5692$\\\hline
\end{tabular}
\caption{\small Slopes of the $1^3S_1$-like trajectories taken
from Ref.\cite{dmli}, All in GeV$^{-2}$.}
\end{center}
\end{table}

 With the help of $\alpha^\prime_{i\bar{j}(N)}=\alpha^\prime_{i\bar{j}(1)}$, from
(\ref{trajectory}) one can have
\begin{eqnarray}
M^2_{i\bar{j}(N)}-M^2_{i\bar{j}(1)}=\frac{\alpha_{i\bar{j}(1)}(0)-\alpha_{i\bar{j}(N)}(0)}{\alpha^\prime_{i\bar{j}(1)}}.
\label{mass-2}
\end{eqnarray}
If $\alpha_{i\bar{j}(1)}(0)-\alpha_{i\bar{j}(N)}(0)$ is simplified to be flavor-independent,
(\ref{mass-2}) can be reduced to the mass formula presented by
Anisovich\cite{anisovich}\footnote{It is expected that trajectories may occur in integrally spaced
sequences, with a `parent' trajectory $\alpha_1(t)$, and an infinite sequence of `daughters'
$\alpha_N(t)=\alpha_1(t)-n_r$, $n_r=0,1,2,3,...$, $N=n_r+1$, see Ref.\cite{regge}. In
Ref.\cite{anisovich}, the mass relation $M^2_N=M^2_0+(N-1)\mu^2,~\mu^2= 1/\alpha^\prime$ is
presented.}
\begin{eqnarray}
M^2_{i\bar{j}(N)}-M^2_{i\bar{j}(1)}=\frac{N-1}{\alpha^\prime_{i\bar{j}(1)}}. \label{mass-3}
\end{eqnarray}
In the presence of $\alpha^\prime_{c\bar{c}(1)}=0.4364$ GeV$^{-2}$
and $M_{J/\psi}=3096.916$ MeV, masses of the $2^3S_1$, $3^3S_1$
and $4^3S_1$ $c\bar{c}$ states predicted by relation
(\ref{mass-3}) are respectively (in MeV) 3447, 3764 and 4058,
which are several hundreds MeV lower than the corresponding
measured masses\cite{pdg} (in MeV) $3686.093\pm 0.034$, $4039\pm
1$ and $4421\pm 4$. This implies the simplification that
$\alpha_{i\bar{j}(1)}(0)-\alpha_{i\bar{j}(N)}(0)$ is
flavor-independent may be too rough for the heavy mesons. The
phenomenological analysis indicates\cite{fili1,fili2} that
$\alpha_{i\bar{j}(1)}(0)-\alpha_{i\bar{j}(N)}(0)$ depends on
masses of the constituent quarks, $m_i$ and $m_j$, and the
functional dependence of
$\alpha_{i\bar{j}(1)}(0)-\alpha_{i\bar{j}(N)}(0)$ on quark masses
is through the combination $m_i+m_j$. Furthermore, the
quantitative results of Ref.\cite{fili1,fili2} show
$\alpha_{i\bar{j}(1)}(0)-\alpha_{i\bar{j}(2)}(0)\approx$ 1.3-1.6.
This idea motivates us to introduce a factor $(1+f_{ij}(m_i+m_j))$
into relation (\ref{mass-3}) for incorporating the corrections due
to the flavor-dependent spacing between $\alpha_{i\bar{j}(1)}(0)$
and $\alpha_{i\bar{j}(N)}(0)$:
\begin{eqnarray}
M^2_{i\bar{j}(N)}-M^2_{i\bar{j}(1)}=\frac{(N-1)}{\alpha^\prime_{i\bar{j}(1)}}(1+f_{ij}(m_i+m_j)),
\label{mass-4}
\end{eqnarray}
where the parameter $f_{ij}$ depends on the flavors $i$ and $j$, and can be obtained by fitting to
the data, masses of the constituent quarks are taken the following values (in GeV)
\begin{eqnarray}
m_u=m_d=0.29, m_s=0.46, m_c=1.65, \label{quarkmass}
\end{eqnarray}
which are typical values used in phenomenological quark
 models\cite{bura1}.

Inserting masses of $J/\psi$ ($1^3S_1$ $c\bar{c}$, mass: 3096.916
MeV), $\psi(4040)$ ($3^3S_1$ $c\bar{c}$, mass: 4039 MeV), $\rho$
($1^3S_1$ $n\bar{n}$, mass: 775.5 MeV), $\rho(1450)$ ($2^3S_1$
$n\bar{n}$, mass: 1459 MeV), $K^\ast(892)$ ($1^3S_1$ $n\bar{s}$,
mass: 896 MeV) and $K^\ast(1580)$ ($2^3S_1$ $n\bar{s}$, mass: 1580
MeV)\footnote{All the masses used as input are taken from
PDG\cite{pdg} except for the $2^3S_1$ kaon mass. The assignment of
the $K^\ast(1410)$ as the $2^3S_1$ kaon is
problematic\cite{k14101,k14102}. Quark model and other
phenomenological approaches consistently suggest the $2^3S_1$ kaon
has a mass about 1580 MeV\cite{mpla}, here we take 1580 MeV as the
mass of $2^3S_1$ $n\bar{s}$.} into the following equations,
\begin{eqnarray}
&&M^2_{\psi(4040)}=M^2_{J/\psi}+(3-1)(1+f_{cc}(m_c+m_c))/\alpha^\prime_{c\bar{c}(1)},\\
&&M^2_{\rho(1450)}=M^2_{\rho}+(2-1)(1+f_{nn}(m_u+m_u))/\alpha^\prime_{n\bar{n}(1)},\\
&&M^2_{K^\ast(1580)}=M^2_{K^\ast(892)}+(2-1)(1+f_{ns}(m_u+m_s))/\alpha^\prime_{n\bar{s}(1)},
\end{eqnarray}
and with the help of the following relations derived from
(\ref{intercept}), (\ref{mass-2}) and (\ref{mass-4})
\begin{eqnarray}
&&2f_{ns}(m_u+m_s)=f_{nn}(m_u+m_u)+f_{ss}(m_s+m_s),\\
&&2f_{cn}(m_c+m_u)=f_{nn}(m_u+m_u)+f_{cc}(m_c+m_c),\\
&&2f_{cs}(m_c+m_s)=f_{cc}(m_c+m_c)+f_{ss}(m_s+m_s),
\end{eqnarray}
one can obtain (in GeV$^{-1}$)
\begin{eqnarray}
f_{nn}=0.601, ~f_{ns}=0.584,~f_{cn}=0.210,~f_{cs}=0.235,~f_{cc}=0.141.
\label{fij}
\end{eqnarray}
Based on the above parameters, we find that
$\alpha_{n\bar{n}(1)}(0)-\alpha_{n\bar{n}(2)}(0)$=1.35,
$\alpha_{n\bar{s}(1)}(0)-\alpha_{n\bar{s}(2)}(0)$=1.44,
$\alpha_{c\bar{n}(1)}(0)-\alpha_{c\bar{n}(2)}(0)$=1.41,
$\alpha_{c\bar{s}(1)}(0)-\alpha_{c\bar{s}(2)}(0)$=1.50, and
$\alpha_{c\bar{c}(1)}(0)-\alpha_{c\bar{c}(2)}(0)$=1.47, which are
in good agreement with the quantitative results given by
Ref.\cite{fili1,fili2} that
$\alpha_{i\bar{j}(1)}(0)-\alpha_{i\bar{j}(2)}(0)\approx$ 1.3-1.6.

 The spectrum of $c\bar{c}$ is well predicted by different theoretical
approaches, and excited vector $c\bar{c}$ states are well
established experimentally, these predicted and measured results
serve us a good testing ground whether our proposed mass relation
(\ref{mass-4}) can give reliable predictions. The measured
$1^3S_1$, $1^3D_1$, $1^3P_0$ and $1^3P_2$ $c\bar{c}$ masses are
used as input\footnote{ The $1^3D_3$ $c\bar{c}$ mass is given by
$\sqrt{\frac{2}{\alpha^\prime_{c\bar{c}(1)}}+M^2_{J/\psi}}$~\cite{dmli}.}
, from (\ref{mass-4}), (\ref{quarkmass}) and (\ref{fij}), our
predicted masses of the radial excitations of the $c\bar{c}$
states lying on the $1^3S_1$-like trajectories are shown in Table
2 and Figure 1. Comparison of results predicted by us and those
from experiments and other theoretical approaches is also given in
Table 2 and Figure 1. Clearly,  masses of the $c\bar{c}$ states
lying on the $1^3S_1$-like trajectories predicted by the mass
relation (\ref{mass-4}) agree remarkably well with those from
measurements and other theoretical approaches.

{\small
\begin{table}[hbt]
\begin{center}
\begin{tabular}{c|ccccccc}\hline
 $c\bar{c}$         & Expt.\cite{pdg} & This work & \cite{p1} & \cite{c1}& \cite{c2}& \cite{ansi} & \cite{gers}  \\\hline
       $1^3S_1$     & 3.097         & {\bf 3.097} & 3.10      & 3.10     & 3.096    & 3.100       & 3.15       \\
       $2^3S_1$     & 3.686           & 3.600     & 3.68      & 3.73     & 3.686    & 3.676       & 3.63          \\
       $3^3S_1$     & 4.039           &{\bf 4.039}     & 4.10      & 4.18     & 4.088    & 4.079       & 4.04              \\
       $4^3S_1$     & 4.421           & 4.434     & 4.45      & 4.56     &          & 4.434       & 4.42              \\

       $1^3D_1$     & 3.771         & {\bf 3.771} & 3.82      & 3.80     & 3.798    & 3.794       & 3.76       \\
       $2^3D_1$     & 4.153           & 4.192     & 4.19      & 4.22     &          & 4.156       & 4.17          \\
       $3^3D_1$     &                 & 4.576     & 4.52      & 4.59     &          & 4.482       & 4.53              \\
       $4^3D_1$     &                 & 4.929     &           &          &          & 4.889       & 4.87              \\
       $1^3P_0$     & 3.415           &{\bf 3.415}     & 3.42      & 3.44     & 3.434    & 3.412       & 3.42          \\
       $2^3P_0$     &                 & 3.875     & 3.92      & 3.94     & 3.854    & 3.867       & 3.86                \\
       $3^3P_0$     &                 & 4.287     &           &          &          & 4.228       & 4.25                \\
       $4^3P_0$     &                 & 4.662     &           &          &          & 4.538       & 4.61                \\

       $1^3D_3$     &                 & 3.765     & 3.85      & 3.83     & 3.815    &             &        \\
       $2^3D_3$     &                 & 4.187     & 4.22      & 4.24     &          &             &           \\
       $3^3D_3$     &                 & 4.571     &           &          &          &             &               \\
       $4^3D_3$     &                 & 4.924     &           &          &          &             &               \\

       $1^3P_2$     & 3.556         & {\bf 3.556} & 3.55      & 3.54     & 3.556    & 3.552       & 3.56          \\
       $2^3P_2$     & 3.929           & 4.000     & 3.98      & 4.02     & 3.972    & 3.986       & 3.98                 \\
       $3^3P_2$     &                 & 4.400     &           &          &          & 4.350       & 4.36                   \\
       $4^3P_2$     &                 & 4.767     &           &          &          & 4.786       & 4.72                   \\
   \hline
\end{tabular}
\caption{\small Masses of $c\bar{c}$ states lying on the
$1^3S_1$-like trajectories. Boldface values stand for masses used
as input. All in GeV.}
\end{center}
\end{table}
}

\begin{figure}[hbt]
\begin{center}
\vspace{-1.5cm} \epsfig{file=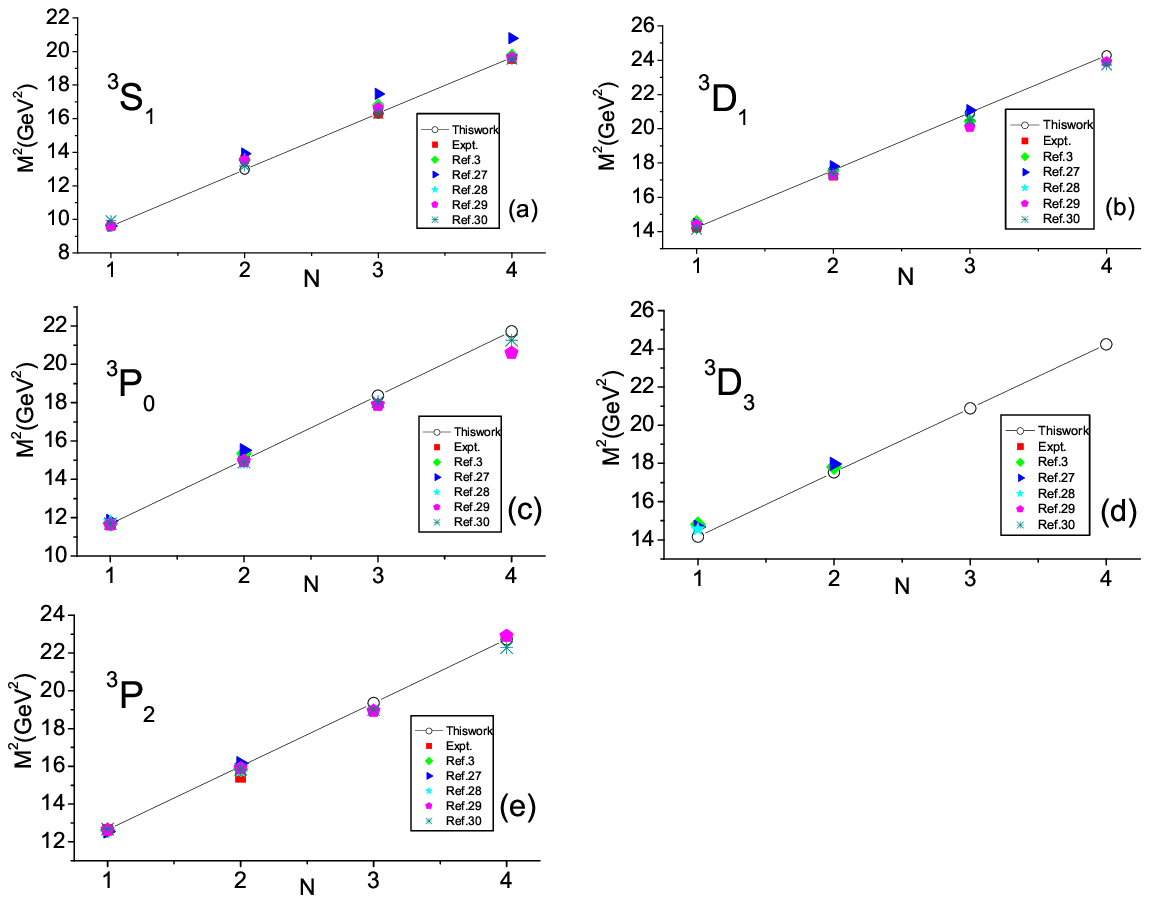,width=12.0cm, clip=}
\vspace*{0.5cm}\vspace*{-1cm}
 \caption{\small The $(N,M^2)$-trajectories for the $c\bar{c}$ states. The predicted masses are displayed numerically in Table 2.}
\end{center}
\end{figure}

\section{Masses of the $c\bar{q}$ states lying on the $1^3S_1$-like trajectories}
\indent \vspace*{-1cm}

In this section, we shall estimate masses of  the ground
$c\bar{q}$ states lying on the $1^3S_1$-like trajectories using
relation (\ref{mass-1}), then the radial excited $c\bar{q}$ masses
can be given by (\ref{mass-4}). In order to derive masses of the
$c\bar{n}$ and $c\bar{s}$ using relation (\ref{mass-1}), we should
know the $n\bar{n}~(I=1)$, $n\bar{s}$ and $c\bar{c}$ masses.
Experimentally, the $n\bar{n} ~(I=1)$, $n\bar{s}$, $c\bar{c}$
states for the $1^3S_1$, $1^3P_2$ and $1^3D_1$ multiplets are well
established\cite{pdg}, however for the $1^3P_0$ $n\bar{n}~(I=1)$,
$n\bar{s}$ and $c\bar{c}$ states, only $c\bar{c}$ state,
$\chi_{c0}(1P)$, is well established, and the assignment for the
$n\bar{n}~(I=1)$ and $n\bar{s}$ remains open. In the recent
literature, there is not yet a consensus on the $1^3P_0$
$n\bar{n}~(I=1)$ mass given by lattice stimulations. For example,
both $M_{a_0(1^3P_0)}\sim 1$ GeV\cite{las1,las2,las3} and
$M_{a_0(1^3P_0)}\sim 1.4-1.6$ GeV\cite{lal1,lal2,lal3} are
predicted recently. In the present work, we shall take
$M_{a_0(1^3P_0)}=(1.0+1.04+1.01)/3\approx 1.02$ GeV, the average
value of predictions given by the recent lattice QCD
calculations\cite{las1,las2,las3} considering that the naive quark
model predicts that the spin-orbit force makes lighter the
$a_0(1^3P_0)$ with respect to the $a_2(1^3P_2)$
($M_{a_2(1^3P_2)}=1.3183$ GeV\cite{pdg}) and the same behavior is
evident in the $c\bar{c}$ and $b\bar{b}$ spectra\cite{k14102}. The
lattice studies suggest the mass of $1^3P_0$ kaon would be 100-130
MeV heavier than the $a_0$ mass \cite{las3}. This is not easily
related to any current experimental candidate, while is consistent
with the result $M_{n\bar{s}(1^3P_0)}=1090\pm 40$ MeV from the K
matrix analysis of the $K\pi$ $S$-wave performed by Anisovich and
Sarantsev\cite{kmatrix}. In this work, we shall take
$M_{n\bar{s}(1^3P_0)}= 1.09$ GeV.

With the help of Table 1 and the following masses (in GeV) used as
input
\begin{eqnarray}
&&M_{n\bar{n}(1^3S_1)}=0.7755,~M_{n\bar{s}(1^3S_1)}=0.896,~
M_{c\bar{c}(1^3S_1)}=3.096916,\nonumber\\
&&M_{n\bar{n}(1^3P_2)}=1.3183,~M_{n\bar{s}(1^3P_2)}=1.4324,~
M_{c\bar{c}(1^3P_2)}=3.5562,\nonumber\\
&&M_{n\bar{n}(1^3P_0)}=1.02,~M_{n\bar{s}(1^3P_0)}=1.09,~
M_{c\bar{c}(1^3P_0)}=3.41476,\nonumber\\
&&M_{n\bar{n}(1^3D_1)}=1.701,~M_{n\bar{s}(1^3D_1)}=1.735,~
M_{c\bar{c}(1^3P_0)}=3.7711,\nonumber
\end{eqnarray}
masses of $c\bar{n}$ and $c\bar{s}$ states predicted by
(\ref{mass-1}) and (\ref{mass-4}) are shown in Table 3 and Figures
2-3, together with the recent predictions by some modified quark
models\cite{matsuki, newmodel1,newmodel2}. The masses of the
$1^3D_3$ $c\bar{n}$ and $c\bar{s}$ are given by\cite{dmli}
\begin{eqnarray}
M_{c\bar{n}(1^3D_3)}=\sqrt{\frac{2}{\alpha^\prime_{c\bar{n}(1)}}+M^2_{c\bar{n}(1^3S_1)}},~M_{c\bar{s}(1^3D_3)}=\sqrt{\frac{2}{\alpha^\prime_{c\bar{s}(1)}}+M^2_{c\bar{s}(1^3S_1)}}.\nonumber
\end{eqnarray}

 {\small
\begin{table}[hbt]
\begin{center}
\begin{tabular}{c|ccccc||c|cccc}\hline
   $D_s$        & Expt.\cite{pdg} & This work & \cite{matsuki}  &\cite{newmodel1}& \cite{newmodel2}   & $D$        & Expt.\cite{pdg}  & This work& \cite{matsuki}   &\cite{newmodel1} \\\hline
       $1^3S_1$ & 2.112           & 2.100     & 2.110           &2.105           & 2.112            &$1^3S_1$      & 2.010            & 2.010      & 2.011          &2.017  \\
       $2^3S_1$ &                 & 2.653     &                 &                & 2.711            &$2^3S_1$      &                  & 2.540      &                &       \\
       $3^3S_1$ &                 & 3.109     &                 &                & 3.153            &$3^3S_1$      &                  & 2.976      &                &        \\
       $1^3D_1$ &                 & 2.775     & 2.817           &                & 2.784            &$1^3D_1$      &                  & 2.738      & 2.762          &     \\
       $2^3D_1$ &                 & 3.214     &                 &                &                  &$2^3D_1$      &                  & 3.147      &                &     \\
       $3^3D_1$ &                 & 3.600     &                 &                &                  &$3^3D_1$      &                  & 3.509      &                &     \\
       $1^3P_0$ & 2.3173          & 2.331     & 2.325           &2.341           & 2.329            &$1^3P_0$      & 2.299$^{\dagger}$& 2.268      & 2.283          &2.260    \\
       $2^3P_0$ &                 & 2.839     &                 &                & 2.817            &$2^3P_0$      &                  & 2.748      &           &         \\
       $3^3P_0$ &                 & 3.269     &                 &                & 3.219            &$3^3P_0$      &                  & 3.156      &                &         \\
       $1^3D_3$ &                 & 2.815     &                 &                &                  &$1^3D_3$      &                  & 2.731      &                &     \\
       $2^3D_3$ &                 & 3.248     &                 &                &                  &$2^3D_3$      &                  & 3.141      &                &     \\
       $3^3D_3$ &                 & 3.630     &                 &                &                  &$3^3D_3$      &                  & 3.504      &                &     \\
       $1^3P_2$ & 2.5735          & 2.562     & 2.568           &2.563           & 2.577            &$1^3P_2$      & 2.459            & 2.457      & 2.468          &2.493    \\
       $2^3P_2$ &                 & 3.032     &                 &                & 3.041            &$2^3P_2$      &                  & 2.906      &           &          \\
       $3^3P_2$ &                 & 3.438     &                 &                &3.431             &$3^3P_2$      &                  & 3.295      &                &          \\
   \hline
\end{tabular}
\caption{\small $D$ and $D_s$ spectra. $^\dagger$Average value of
2290 and 2308 MeV by Belle. All in GeV.}
\end{center}
\end{table}
 }
\begin{figure}[hbt]
\begin{center}
\vspace{-1.5cm} \epsfig{file=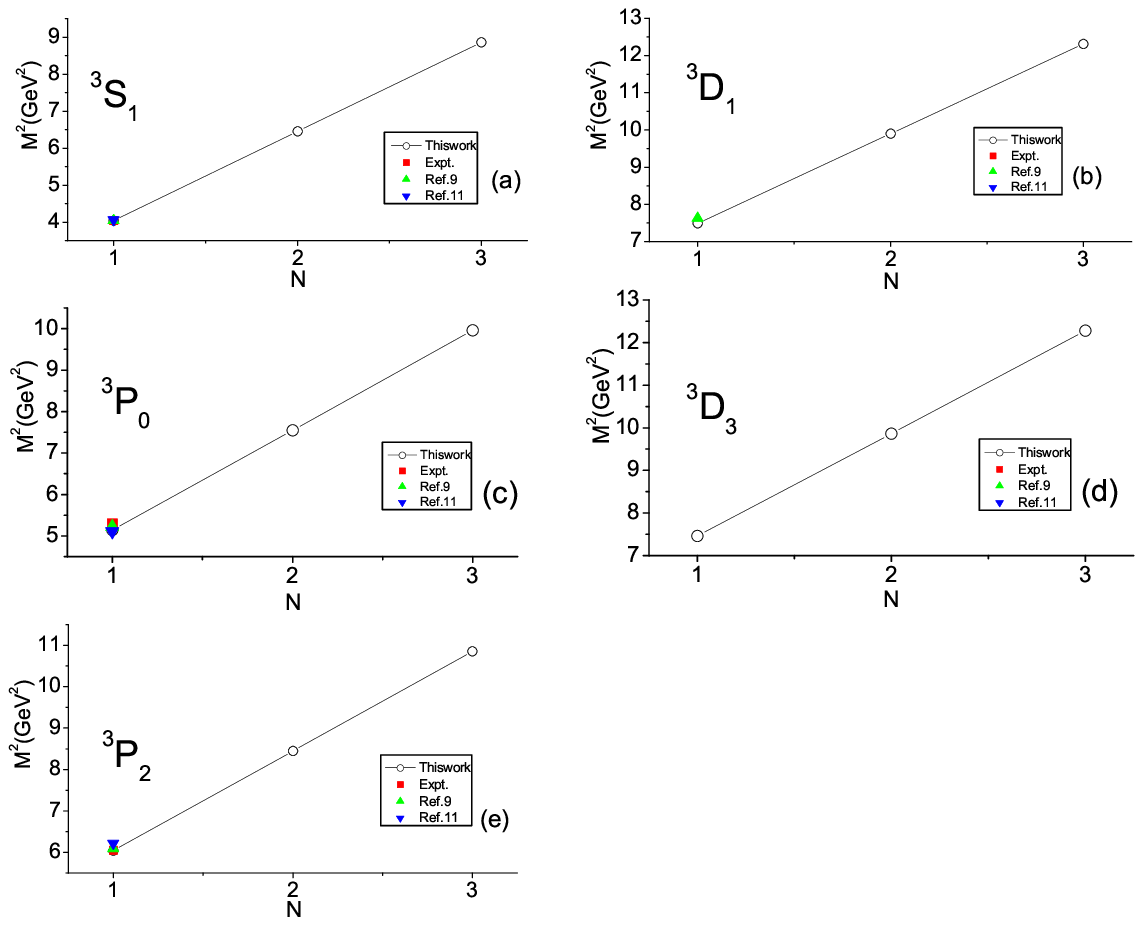,width=12.0cm, clip=}
\vspace*{0.5cm}\vspace*{-1cm}
 \caption{\small The $(N,M^2)$-trajectories for the $c\bar{n}$ states. The predicted masses are displayed numerically in Table 3.}
\end{center}
\end{figure}

\begin{figure}[htb]
\begin{center}
\vspace{-1.5cm} \epsfig{file=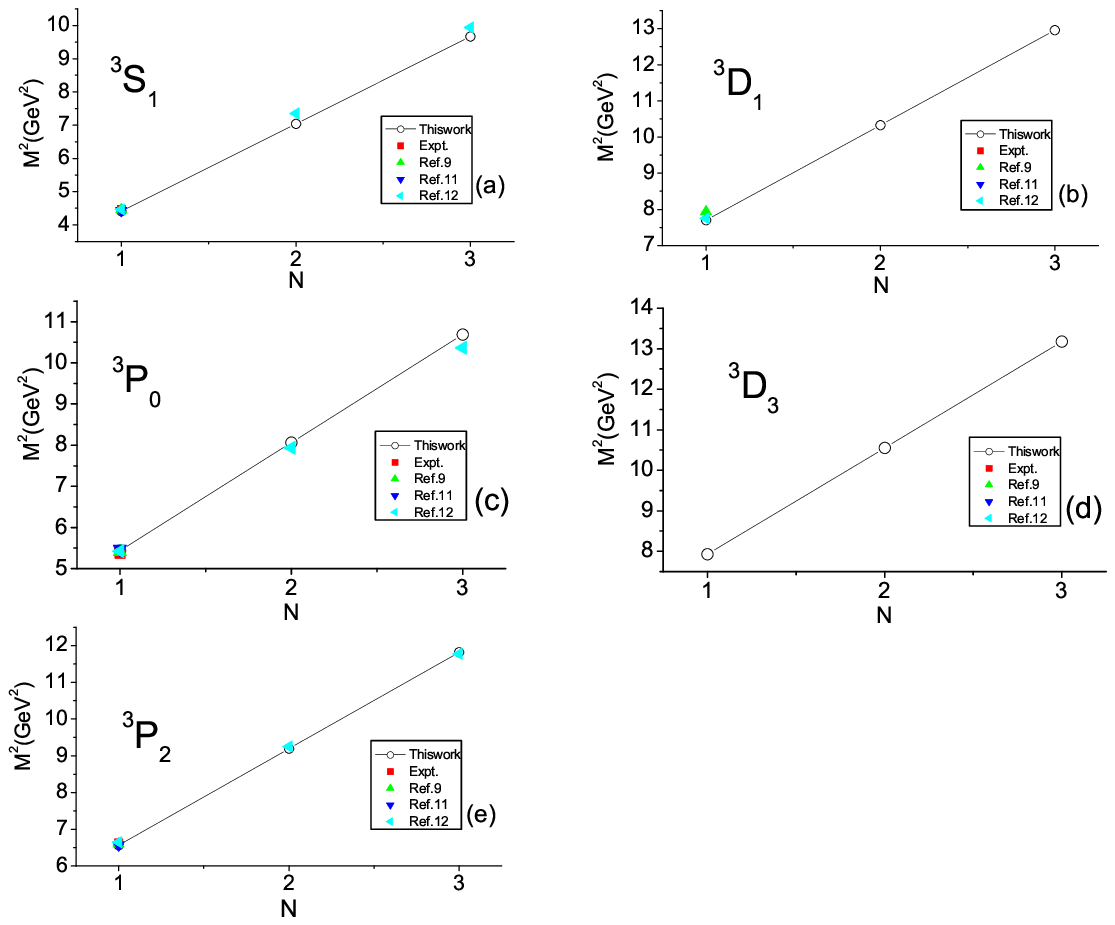,width=12.0cm, clip=}
\vspace*{0.5cm}\vspace*{-1cm}
 \caption{\small The $(N,M^2)$-trajectories for the $c\bar{s}$ states. The predicted masses are displayed numerically in Table 2.}
\end{center}
\end{figure}

\section{Discussions}
\indent\vspace*{-1cm}

From Table 3 and Figures 2-3, it is clear that the agreement
between our predicted $D$ and $D_{s}$ masses and the recent
predictions by some modified quark
model\cite{matsuki,newmodel1,newmodel2} is quite good. It
therefore appears likely that Regge phenomenology is capable of
describing the $D$ and $D_{s}$ masses with reasonable accuracy.

Our predicted $D_0$ meson mass is 2.268 GeV, in good agreement
with the preliminary Belle measurement of $2290\pm 22\pm 20$
MeV\cite{Belle1} and the current Belle mass of $2308\pm 17\pm 32$
MeV\cite{Belle2} within one to two percent of accuracy, while in
disagreement with the FOCUS mass of $2407\pm 21\pm 35$
MeV\cite{focus}. For the $D_{s0}$ mass, our prediction is 2.331
GeV, only 13.7 MeV higher than the experimental result, and in
good agreement with the recent result predicted by QCD sum rule
that $M_{c\bar{s}(1^3P_0)}=2.31\pm 0.03$ GeV\cite{0610327}. Our
analysis supports the conclusion that the $D_{s0}(2317)$ can be
identified as a conventional $1^3P_0$ $c\bar{s}$ state.

Based on the measured masses of
 $n\bar{n}$, $n\bar{s}$ and $c\bar{c}$ states for the $1^3S_1$
and $1^3P_2$ multiplets, the predicted masses for the $1^3S_1$ and
$1^3P_2$ $c\bar{q}$ by relation (\ref{mass-1}) are in excellent
agreement with the measurements. In the presence of
$M_{n\bar{n}(1^3P_0)}=1.02$ GeV and $M_{n\bar{s}(1^3P_0)}=1.09$
GeV, the relation (\ref{mass-1}) gives an accurate determination
for the $D_0$ and $D_{s0}$ masses, it implies that the $a_0(980)$
could be indeed the lightest non-singlet scalar meson.

BaBar recently reported the discovery of a new $D_{sJ}(2860)$
state with a mass of $2856.6\pm 1.5\pm 5.0$ MeV and width of
$48\pm 7\pm 10$ MeV\cite{0607082}. BaBar observed this state only
in the $D^0K^+$ or $D^+K^0_S$ system and found no evidence in the
$D^{\ast 0}K^+$ and $D^{\ast +}K^0_S$ channels, therefore the
possible $J^{P}$ of the $D_{sJ}(2860)$ is $0^+,~
1^-,~2^+,~3^-,\cdots$.

Based on our prediction as shown in Table 3, the $D_{sJ}(2860)$
mass is about 213 MeV higher than the $2^3S_1$ $c\bar{s}$ mass
(2653 MeV), and 82 MeV higher than the $1^3D_1$ $c\bar{s}$ mass
(2775 MeV).
 The decay
pattern and total width of the $D_{sJ}(2860)$ for various quantum
number assignments has been investigated by Zhang et
al.\cite{0609013} in a $^3P_0$ decay model. If the $D_{sJ}(2860)$
is the $2^3S_1$ $c\bar{s}$ state, its total width is about 90 MeV,
and the dominant decay modes are $DK^\ast$ with
$\Gamma(DK^\ast)=24$ MeV and $D^\ast K$ with $\Gamma(D^\ast K)=64$
MeV, and the $DK$ mode is only 12 KeV\cite{0609013}. If the
$D_{sJ}(2860)$ is the $1^3D_1$ $c\bar{s}$ state, its total width
is about 132 MeV, and $\Gamma(DK):\Gamma(D_s\eta):\Gamma(D^\ast
K):\Gamma(D^\ast_s\eta):\Gamma(DK^\ast)\approx
42:12:7:1:4$\cite{0609013}. Both the predicted mass and width for
the $2^3S_1$ or $1^3D_1$ $c\bar{s}$ are significantly larger than
the experimental data of the $D_{sJ}(2860)$, making the assignment
of the $D_{sJ}(2860)$ as the $2^3S_1$ or $1^3D_1$ $c\bar{s}$
unfavorable.

However, the $D_{sJ}(2860)$ mass is close to our predicted
$2^3P_0~c\bar{s}$ mass (2839 MeV) and $1^3D_3~c\bar{s}$ mass (2815
MeV). If the $D_{sJ}(2860)$ is the $2^3P_0$ $c\bar{s}$ state, its
total width becomes around 54 MeV, consistent with the measured
width of the $D_{sJ}(2860)$, $48\pm 7\pm 10$ MeV, within errors,
the dominant decay modes are $DK$ with $\Gamma(DK)=37$ MeV and
$D_s\eta$ with $\Gamma(D_s\eta)=16$ MeV\cite{0609013}, and the
decay modes $D^\ast K$, $D^\ast_s\eta$ and $DK^\ast$ are
forbidden. The suggestion that the $D_{sJ}(2860)$ can be
identified as the $2^3P_0$ $c\bar{s}$ has been given by
\cite{newmodel2,ex1}. If the $D_{sJ}(2860)$ is the $1^3D_3$
$c\bar{s}$ state, its total width is about 37 MeV, also compatible
with $48\pm 7\pm 10$ MeV within errors, the dominant decay modes
are $DK$ with $\Gamma(DK)=22$ MeV and $D^\ast K$ with
$\Gamma(D^\ast K)=13$ MeV\cite{0609013}. The assignment of the
$D_{sJ}(2860)$ as the $3^-$ $c\bar{s}$ has been proposed by
\cite{ex2}. At present, only the decay $D_{sJ}(2860)\rightarrow
DK$ is observed experimentally, which is not enough to distinguish
the above two possible assignments, therefore, the assignment of
the $D_{sJ}(2860)$ as the $2^3P_0$ or $1^3D_3$ $c\bar{s}$ seems
favorable by the available experimental information.

It should be noted that for the $2^3P_0$ $c\bar{s}$ state, the
decay modes $D^\ast K$, $D^\ast_s\eta$ and $DK^\ast$ are
forbidden, and the $1^3D_3$ $c\bar{s}$ state has a large $D^\ast
K$ decay width and a small $D_s\eta$ decay width ($\sim 1.2$
MeV\cite{0609013}), therefore, the further experimental search of
the $D_{sJ}(2860)$ in the $DK^\ast$, $D^\ast K$, $D^\ast_s\eta$
and $D_s\eta$ decay modes would be certainly desirable for
distinguish the above two possible assignments.

Our predicted mass of the $2^3S_1$ $c\bar{s}$ state is around 2653
MeV, in excellent agreement with the prediction $2658\pm 15$ MeV
obtained by Chang from Bethe-Salpeter equation\cite{chang}, and
48~MeV lighter than the $D_{sJ}(2690)/D_{sJ}(2700)$
mass\footnote{The average value of the BaBar mass and Belle mass
is~2701 MeV}. If the $D_{sJ}(2690)/D_{sJ}(2700)$ is the $2^3S_1$
$c\bar{s}$ state, its total width is about 103 MeV in a $^3P_0$
decay model\cite{newmodel2}, consistent with the measured width of
$112\pm 7\pm 36$ or $115\pm 20^{+36}_{-32}$ MeV, and the dominant
decay modes are $DK$ with a width of 22 MeV and $D^\ast K$ with a
width of 78 MeV\cite{newmodel2}. Also, the
$D_{sJ}(2690)/D_{sJ}(2700)$ mass is about 74 MeV lighter than our
predicted $1^3D_1~c\bar{s}$ mass (2775 MeV), if it is the
$1^3D_1~c\bar{s}$ state, the calculations performed by
Ref.\cite{0609013} within a $^3P_0$ model show its total width is
73 MeV, also roughly consistent with the experimental data, and
the dominant decay modes are $DK$ with a width of about 49 MeV and
$D_s\eta$ with a width of about 13 MeV. Therefore, the
$D_{sJ}(2690)/D_{sJ}(2700)$ as either a $2^3S_1$ or $1^3D_1$
$c\bar{s}$ seems consistent with experimental data\cite{0609013}.
Considering that the $D_{sJ}(2690)/D_{sJ}(2700)$ mass is about 48
MeV higher than $M_{c\bar{s}(2^3S_1)}$, while 74 MeV lower than
$M_{c\bar{s}(1^3D_1)}$, we tend to suggest the
$D_{sJ}(2690)/D_{sJ}(2700)$ is most likely a mixture of the
$D_s(2^3S_1)$ and $D_s(1^3D_1)$. It has been found
that\cite{newmodel2} in the presence of the $2^3S_1$ $c\bar{s}$
mixing with $1^3D_1$ $c\bar{s}$ (with the mixing angle of
approximately $-0.5$ radians), the predicted total width of the
$D_{sJ}(2690)/D_{sJ}(2700)$ (with mass set to 2688 MeV) by a
$^3P_0$ model becomes about 110 MeV, in good agreement with the
data.

\section{Summary and conclusion}
\indent\vspace*{-1cm}

Masses of the $D$ and $D_s$ states lying on the $1^3S_1$-like
trajectories are estimated in Regge phenomenology. The predicted
masses agree well with the recent results by some modified quark
models. It therefore appears likely that Regge phenomenology is
capable of describing the $D$ and $D_{s}$ masses with reasonable
accuracy.

Our predictions show that masses of the recent observed charmed
states such as the $D_{s0}(2317)$, $D_{sJ}(2860)$ and
$D_{sJ}(2690)/D_{sJ}(2700)$ can be reasonably reproduced in the
simple quark-antiquark picture. Based on our analysis, we suggest
the $D_{sJ}(2317)$ can be identified as the conventional $1^3P_0$
$c\bar{s}$ state, the assignment of the $D_{sJ}(2860)$ as the
$D_s(2^3P_0)$ or $D_s(1^3D_3)$ seems favorable, and the
$D_{sJ}(2690)/D_{sJ}(2700)$ is most likely a mixture of the
$D_s(2^3S_1)$ and $D_s(1^3D_1)$.

\noindent {\bf Acknowledgments:}
 This work
is supported in part by National Natural Science Foundation of
China under Contract No. 10205012, Henan Provincial Science
Foundation for Outstanding Young Scholar under Contract No.
0412000300, Program for New Century Excellent Talents in
University of Henan Province under Contract No. 2006HANCET-02, and
Program for Youthful Excellent Teachers in University of Henan
Province.


\baselineskip 18pt
\begin{thebibliography}{99}
\bibitem{ds2317ex}BaBar Collaboration, Phys. Rev. Lett. {\bf 90},
242001 (2003); CLEO Collaboration, Phys. Rev. D {\bf 68}, 032002
(2003); Belle Collaboration, Phys. Rev. Lett. {\bf 92}, 012002
(2004); FOCUS Collaboration, hep-ph/0406044
\bibitem{pdg} W. M. Yao et al., J. Phys. G {\bf 33}, 1 (2006)
\bibitem{p1}S. Godfrey, N. Isgur, Phys. Rev. D {\bf 32}, 189 (1985)
\bibitem{p2}M. Di Pierro, E. J. Eichten, Phys. Rev. D {\bf 64}, 114004 (2001)
\bibitem{rmp} S. Godfrey and J. Napolitano, Rev. Mod. Phys. {\bf 71}, 1411 (1999)
\bibitem{rev1} W. Lucha, F. F. Schoberl, Mod. Phys. Lett. A {\bf
18}, 2837 (2003);
\bibitem{rev2}P. Colangelo, F. De Fazio and R. Ferrandes,
Mod. Phys. Lett. A {\bf 19}, 2083 (2004); R. Ferrandes,
hep-ph/0407212
\bibitem{rev3} E. S. Swanson, Phys. Rept. {\bf 429}, 243 (2006)
\bibitem{matsuki} T. Matsuki, T. Morii and K. Sudoh, hep-ph/0605019
\bibitem{loop}Ian Woo Lee et al.,  hep-ph/0412210
\bibitem{newmodel1}O. Lakhina and S. Swanson, hep-ph/0608011
\bibitem{newmodel2}F. E. Close, C. E. Thomas, O. Lakhina and E. S. Swanson, hep-ph/0608139
\bibitem{0607082} BaBar Collaboration, hep-ex/0607082
\bibitem{0608031} Belle Collaboration, hep-ex/0608031

\bibitem{regge} P. D. Collins, An introduction to
Regge theory and high energy physics, Cambridge University Press,
1977
\bibitem{bas} L. Basdevant, S. Boukraa, Z. Phys. C {\bf  28}, 413 (1985)
\bibitem{anisovich} A. V. Anisovich, V. V. Anisovich and A. V. Sarantsev, Phys. Rev. D {\bf  62}, 051502 (2000); V. V. Anisovich, hep-ph/0110326; {\sl ibid.} hep-ph/0208123; {\sl ibid.} hep-ph/0310165
\bibitem{nonlinear1} W. K. Tang, Phys. Rev. D {\bf  48}, 2019 (1993); A. Brandat et al., Nucl. Phys. B {\bf  514}, 3 (1999);
M. M. Brisudova, L. Burakovsky and T. Goldmann, Phys. Lett. B {\bf
460}, 1 (1999);
  A. Tang, J. W. Norbury, Phys. Rev. D {\bf  62}, 016006 (2000); L. Pando Zayas, J. Sonnenschein, D. Vaman, hep-th/0311190; S. S. Afonin et al., hep-ph/0403268
\bibitem{2} M. M. Brisudova, L. Burakovsky and T. Goldmann,
Phys. Rev. D {\bf 61}, 054013 (2000)
\bibitem{dmli} De-Min Li et al.,  Eur. Phys. J. C {\bf 37}, 323 (2004)
\bibitem{fili1} S. Filipponi and Y. Srivastava, Phys. Rev. D {\bf
58}, 016003 (1998)
\bibitem{fili2}S. Filipponi, G. Pancheri and Y. Srivastava, Phys.
Rev. Lett. {\bf 80}, 1838 (1998)
\bibitem{bura1} L. Burakovsky, T. Goldman, Phys. Lett. B {\bf 434}, 251
(1998); Phys. Rev. Lett. {\bf 82}, 457 (1999)
\bibitem{k14101} T. Barnes, N. Black and P. R. Page,  Phys. Rev. D
{\bf 68}, 054014 (2003)
\bibitem{k14102} J. Vijande, F. Fernandez
and A. Valcarce, J. Phys. G {\bf 31}, 481 (2005)
\bibitem{mpla} De-Min Li et al., Mod. Phys. Lett. A {\bf 20}, 2497
(2005)
\bibitem{c1}J. Zeng, J. V. Van Orden and W. Roberts, Phys. Rev. D {\bf 52},5229 (1995)
\bibitem{c2} D. Ebert, R. N. Faustov and V. O. Galkin, Phys. Rev. D {\bf 67}, 014027 (2003)
\bibitem{ansi} V. V. Anisovich et al., hep-ph/0511005
\bibitem{gers} S. S. Gershtein, A. K. Likhoded and A. V.
Luchinsky, Phys.Rev. D {\bf 74}, 016002 (2006)
\bibitem{las1}S. Prelovsek and K. Orginos, Nucl. Phys. Proc. Suppl. {\bf 119}, 822
(2003)
\bibitem{las2}A. Hart, C. McNeile and C. Michael, Nucl. Phys. Proc. Suppl.
{\bf 119}, 266 (2003)
\bibitem{las3}C. McNeile and C. Michael, Phys. Rev. D {\bf 74}, 014508
(2006)
\bibitem{lal1} S. Prelovsek et al, Phys. Rev. D {\bf 70}, 094503
(2004)
\bibitem{lal2} T. Burch et al., Phys.Rev. D {\bf 73},
094505 (2006)
\bibitem{lal3} N. Mathur et al., hep-ph/0607110
\bibitem{kmatrix} A. V. Anisovich and A. V. Sarantsev, Phys. Lett.
B {\bf 413}, 137 (1997)
\bibitem{Belle1}Belle Collaboration, BELLE-CONF-0235, ICHEP02
abstract 724 (2002)

\bibitem{Belle2} Belle Collaboration, Phys. Rev. D {\bf 69},
112002 (2004)
\bibitem{focus} FOCUS Collaboration, Phys. Lett. B {\bf 586}, 11
(2004)
\bibitem{0610327} Y. B. Dai, S. L. Zhu and Y. B. Zuo,
hep-ph/0610327
\bibitem{0609013} B. Zhang, X. Liu, W. Z. Deng and S. L. Zhu, hep-ph/0609013
\bibitem{ex1}E. V. Beveren ang G. Rupp, Phys. Rev. Lett. {\bf 97}, 202001
(2006)
\bibitem{ex2} P. Colangelo, F. D. Fazio and S. Nicotri,
Phys. Lett. B {\bf 642}, 48 (2006)
\bibitem{chang} C. H. Chang, C. S. Kim and G. L. Wang, Phys. Lett.
B {\bf 623}, 218 (2005)
\end{thebibliography}
\end{document}